\documentclass[useAMS,usegraphicx]{mn2e}

\def\gsim{\mathrel{\raise.5ex\hbox{$>$}\mkern-14mu
             \lower0.6ex\hbox{$\sim$}}}
\def\lsim{\mathrel{\raise.3ex\hbox{$<$}\mkern-14mu
             \lower0.6ex\hbox{$\sim$}}}

\title[The Cosmic Battery and the ISCO]{The Cosmic Battery
and the Inner Edge\\ of the Accretion Disk}
\author[I. Contopoulos and D. B. Papadopoulos]{I. Contopoulos$^{1}$\thanks{E-mail:
icontop@academyofathens.gr (IC)} and D. B.
Papadopoulos$^{2}$\\
$^{1}$Research Center for Astronomy and Applied
Mathematics, Academy of Athens, Athens 11527, Greece\\
$^{2}$Department of Physics, Aristotle University of Thessaloniki,
Thessaloniki 54124, Greece}
\begin{document}

\date{Accepted . Received ; in original form}

\pagerange{\pageref{firstpage}--\pageref{lastpage}}
\pubyear{20012}

\maketitle

\label{firstpage}

\begin{abstract}
The Poynting-Robertson Cosmic Battery proposes that the innermost
part of the accretion disk around a black hole is threaded by a
large scale dipolar magnetic field generated in situ, and that the
return part of the field diffuses outward through the accretion
disk. This is different from the scenario that the field
originates at large distances and is carried inward by the
accretion flow. In view of the importance of large scale magnetic
fields in regulating the processes of accretion and outflows, we
study the stability of the inner edge of a magnetized disk in
general relativity when the distribution of the magnetic field is
the one predicted by the Poynting-Robertson Cosmic Battery. We
found that as the field grows, the inner edge of the disk
gradually moves outward. In a fast spinning black hole with
$a\gsim 0.8M$ the inner edge moves back in towards the black hole
horizon as the field grows beyond some threshold value. In all
cases, the inner part of the disk undergoes a dramatic structural
change as the field approaches equipartition.
\end{abstract}


\section{Introduction}

Accretion of matter onto astrophysical black holes is a complex
astrophysical process involving several physical sub-processes
(general relativity, small and large scale magnetic fields,
turbulence, plasma microphysics, etc.) Accretion proceeds in the
form of a disk in which material moves slowly inward until it
reaches an inner edge, beyond which it plunges in on dynamical
timescales. Understanding where the accretion disk ends is of
paramount importance in the two methods for measuring black hole
spin, namely fitting the thermal X-ray continuum (Zhang, Cui \&
Chen~1997), and fitting the profile of the relativistically
broadened iron K$\alpha$ line (Fabian {\em et al.}~1989; see also
McClintock et al.~2011 and references therein).

The inner edge of the disk is expected to be close to the radius
of the Innermost Stable Circular Orbit (hereafter ISCO). It is
well known that, for test particles, the position of the ISCO
depends on the mass and angular momentum of the black hole. The
situation is more complicated when, instead of test particles, one
considers plasma in orbit around the black hole. It has been shown
numerically that any turbulent magnetic field that may thread the
plasma modifies its dynamics with respect to the idealized case of
test particles, and the inner edge of the disk moves inward (Penna
{\em et al.}~2010). It has even been suggested that since the bulk
of the disk is turbulent, there is no feature in the flow related
to the ISCO, and the inner edge of the disk moves inward very
close to the event horizon (Balbus~2012). What is still not very
clear is what is the role of any large scale dipolar magnetic
field present in the immediate vicinity of the central black hole.
Igumenshchev~(2008), Tchekhovskoy, Narayan \& McKinney~(2011) and
others have shown numerically that a dipolar magnetic field
advected inward from large distances can grow to equipartition,
and can disrupt the process of accretion forming a `magnetically
arrested disk' (Narayan, Igumenshchev \& Abramowicz~2003)
interspersed with inward spiraling bundles of low-density plasma.

An independent line of thought, the Poynting-Robertson Cosmic
Battery (hereafter PRCB; Contopoulos \& Kazanas~1998; Kylafis,
Contopoulos, Kazanas \& Christodoulou~2011), proposes that the
central dipolar magnetic field is generated in situ, with the
return part of the field diffusing outward through the turbulent
accretion disk. In other words, the origin of the large scale
magnetic field is situated in the vicinity of the inner edge of
the accretion disk, and not at large distances as in the latter
case. In view of the importance of large scale magnetic fields in
regulating the processes of accretion and outflows, we believe
that there is still a lot to be learned from the theoretical study
of axisymmetric accretion within the framework of the PRCB. We,
therefore, decided to follow on the footsteps of Thorne \&
Macdonald~(1982) and Mobarry \& Lovelace~(1986) and determine the
inner edge of a magnetized disk in general relativity when the
distribution of the large scale magnetic field is the one
predicted by the PRCB. For simplicity, we ignore here the
potentially important role of turbulent magnetic stresses
(Balbus~2012). We begin in the next section with a presentation of
our equations and physical simplifications of the problem. In \S~3
we obtain our first results for a non-rotating (Schwarzschild)
black hole, and in \S~4 we extend our study in the most general
case of a rotating (Kerr) black hole. Our conclusions are drawn in
\S~5.

\section{The Relativistic MHD Equations}

Throughout this paper we adopt geometric units where $G=c=1$.
Semicolon stands for covariant derivative, comma for partial
derivative. Latin indices denote spatial components $(1-3)$, Greek
indices denote space-time components $(0-3)$, and `$\sim $'
denotes the spatial part of vectors. In order to derive the
relativistic MHD equations for our problem, we follow closely the
3+1 formulation of Thorne \& Macdonald~(1982) used by most
researchers in the astrophysical community. We start with the most
general 4-dimensional space-time geometry
\begin{equation}\label{st1}
{\rm d}s^2=g_{\mu\nu}{\rm d}x^\mu {\rm d}x^\nu\ ,
\end{equation}
and the Einstein field equations
\begin{equation}\label{e1}
R_{\mu\nu}-\frac{1}{2}g_{\mu\nu}R=8\pi T_{\mu\nu}
\end{equation}
with the Bianchi identities
\begin{equation}\label{eq2}
(R^{\mu\nu}-\frac{1}{2}g^{\mu\nu}R)_{;\mu\nu}=0\ ,
\end{equation}
where $T^{\mu\nu}$, the energy-momentum tensor, satisfies
\begin{equation}\label{e3}
T^{\mu\nu}_{;\nu}=0\ .
\end{equation}
We will restrict our analysis to space times where the torsion and
expansion (but not the shear) vanish. In a cylindrical system of
coordinates $x^\mu\equiv (t,r,\theta,\phi)$ we henceforth also
assume {\em steady-state}, $(\ldots)_{,t}=0$, and {\em
axisymmetry}, $(\ldots)_{,\phi}=0$. The general 4-dimensional
space-time geometry of Eq.~(\ref{st1}) may be rewritten as
\begin{equation}\label{st2}
{\rm d}s^2=-\alpha^2{\rm d}t^2+g_{\phi\phi}({\rm d}\phi-\omega{\rm
d}t)^2+g_{rr}{\rm d}^2r+g_{\theta\theta}{\rm d}\theta^2\ .
\end{equation}
Notice that the theory of steady axisymmetric flows in ideal
General Relativistic Magneto-Hydro-Dynamics (hereafter GRMHD)
around a Schwarzschild black hole in (3+1) space-time split
formulation was first developed by Mobarry \& Lovelace~(1986). We
choose fiducial {\it `zero-angular momentum observers'} or ZAMOs
who consider hypersurfaces $t=$~constant as `slices of
simultaneity'. ZAMOs move with 4-velocity
\begin{equation} \label{ZAMO}
U^\mu=\frac{1}{\alpha}(1\ ,0\ ,0\ ,\omega)
\end{equation}
orthogonal to hypersurfaces of constant $t$. They are non-inertial
observers with acceleration
\begin{equation}\label{acceleration}
{\rm a}^\mu \equiv U^\mu_{;\nu}U^\nu\ .
\end{equation}
Here, the lapse function of our observers $\alpha\equiv
(g^2_{t\phi}/g_{\phi\phi}-g_{tt})^{1/2}$, $\omega\equiv
-g_{t\phi}/g_{\phi\phi}$ and $g_{ii}$ are in general functions of
$r$ and $\theta$. Finally, we define the tensor that projects
4-vectors onto hypersurfaces of constant $t$ as
\begin{equation}\label{gamma}
\gamma^{\mu\nu}\equiv g^{\mu\nu}+U^\mu U^\nu\ .
\end{equation}
Our hydromagnetic system will be specified by the following choice
for the energy-momentum tensor
\begin{equation}\label{e4}
T^{\mu\nu}=\rho u^{\mu}u^{\nu}
+\frac{1}{4\pi}(F_{\alpha}^{\mu}
F^{\nu\alpha}-\frac{1}{4}F_{\alpha\beta} F^{\alpha\beta}
g^{\mu\nu})\ .
\end{equation}
Here, $u^{\mu}\equiv {\rm d}x^\mu/{\rm d}\tau$ is the fluid
4-velocity as measured by ZAMOs, and $\rho$ is the density of the
fluid. Notice that we have set here the pressure equal to zero,
i.e. we have assumed a cold plasma. $F^{\mu \nu}$ is the
electromagnetic field tensor which is related to the electric and
magnetic fields $E^\mu, B^\mu$ measured by ZAMOs through
\begin{equation}
F^{\mu \nu}=U^\mu E^\nu -U^\nu E^\mu +
\epsilon^{\mu\nu\lambda\rho}B_\lambda U_\rho\ . \label{Fmunu}
\end{equation}
Here, $\epsilon_{\mu\nu\lambda\rho}\equiv \sqrt{|{\rm det}(g_{\mu
\nu})|}[\mu \nu \lambda \rho]$ is the 4-dimensional Levi-Civita
tensor. The GRMHD equations of motion are obtained from
Eq.(\ref{e3}) supplemented by Maxwell's equations of
electrodynamics
\begin{equation}\label{eq5}
F^{\mu\nu}_{;\nu}=4\pi J^\mu,~~F_{[\mu\nu;\alpha]}=0\ ,
\end{equation}
where $J^{\mu}$ is the electric current density that satisfies the
charge conservation condition $J^{\mu}_{;\mu}=0$. The electric
current is related to the electric field through a generalized
Ohm's law. Eqs.~(\ref{e3}) and (\ref{eq5}) now read
\[
\tilde{\nabla}\cdot \tilde{B} =0\nonumber
\]
\[
\frac{1}{\alpha}\tilde{\nabla}\cdot (\alpha \tilde{J}) =0\nonumber
\]
\[
D_{\tau}\tilde{B} =0\nonumber
\]
\[
D_{\tau}\varepsilon+\frac{1}{\alpha^2}\tilde{\nabla}
\cdot(\alpha^2 \tilde{S})+W^{ij}\sigma_{ij} =0\nonumber
\]
\begin{equation}
D_{\tau}\tilde{S}+\tilde{\sigma}\cdot \tilde{S}+\varepsilon
\tilde{\rm a}+\frac{1}{\alpha}\tilde{\nabla}\cdot(\alpha
\tilde{W})
=0\ ,\label{k21}
\end{equation}
where,
\[
\varepsilon =
\Gamma^2\rho+\frac{1}{8\pi}(\tilde{E}^2+\tilde{B}^2)\ ,\ \
\tilde{S} = \Gamma^2\rho \tilde{\rm
v}+\frac{1}{4\pi}\tilde{E}\times \tilde{B}\ ,\nonumber
\]
\[
 \tilde{W} = \Gamma^2\rho \tilde{\rm v}\otimes\tilde{\rm v}
 +\frac{1}{4\pi}\left[-(\tilde{E}\otimes\tilde{E}
 +\tilde{B}\otimes\tilde{B})+\frac{1}{2}(\tilde{E}^2
 +\tilde{B}^2)\tilde{\gamma}\right]\ ,\nonumber
 \]
 \[
\sigma^{ij}
 =\frac{1}{2}\gamma^{i\mu}\gamma^{j\nu}(U_{\mu ; \nu}+
 U_{\nu ; \mu})\ ,\ \ \tilde{J}=\frac{\tilde{\nabla}\times (\alpha
\tilde{B})}{4\pi\alpha}\ ,\ \ \mbox{and}\nonumber
\]
\[
\Gamma \equiv (1-\tilde{\rm v}^2)^{-1/2}\equiv
(1+\tilde{u}^2)^{1/2}\ .
\]
Here,
$\epsilon$ is the internal energy, $\sigma^{\mu \nu}$ is the shear
tensor of our system of reference, $\tilde{\rm v}\equiv
\alpha^{-1}{\rm d}x^i/{\rm d}t$ is the fluid 3-velocity as
measured by ZAMOs ($\tilde{\rm v}\equiv \Gamma^{-1}\tilde{u}$),
and $D_\tau$ is the Fermi derivative (see Appendix for details).
We simplify the problem even further by assuming a dipolar
magnetic field geometry, by restricting our analysis to the
equatorial plane $\theta=\pi/2$, and by considering only the
simplest case with no azimuthal magnetic torques. In
$(r,\theta,\phi)$ coordinates, symmetry dictates that
\begin{eqnarray}\label{symmetry}
\tilde{\rm v} & = & (0,0,{\rm v}^\phi)\nonumber\\
\tilde{B} & = & (0,B^\theta,0)\ \ \ \mbox{with}\ \ \
B^r_{,\theta}\neq 0\ \
\ \mbox{in general}\ ,\nonumber\\
\tilde{E} & = & (E^r, 0, 0)\ \ \ \mbox{with}\ \ \
E^\theta_{,\theta}\neq 0\ \ \ \mbox{in general}\ .
\end{eqnarray}
For several applications in astrophysics, perfect (infinite)
conductivity is a valid approximation. In that limit, Ohm's law
yields the ideal MHD condition
\begin{equation}\label{ideal}
u_{\mu} F^{\mu\nu}=0
\end{equation}
which is equivalent to the statement that the electric field
vanishes in a frame comoving with the fluid.

\section{The Schwarzschild metric}

Before addressing the general case of an accreting spinning black
hole, we will first consider the Schwarzschild metric on the
equatorial plane where
\begin{eqnarray}\label{k1}
{\rm d}s^2 & = & g_{tt} {\rm d}t^2+g_{rr} {\rm
d}r^2+g_{\theta\theta} {\rm d}\theta^2
+g_{\phi\phi} {\rm d}\phi^2\nonumber\\
& = & -(1-\frac{2M}{r}){\rm d}t^2+(1-\frac{2M}{r})^{-1} {\rm
d}r^2\nonumber\\
 & & +r^2 {\rm d}\theta^2+r^2{\rm d}\phi^2\ .
\end{eqnarray}
Here, $M$ is the mass of the black hole. The lapse function is
equal to $\alpha=(1-2M/r)^{1/2}$. $\omega=0$ and $\sigma^{\mu
\nu}=0$. Furthermore,
\begin{eqnarray}
U^\mu & = & \frac{1}{\alpha}\left(1\ ,0\ ,0\
,0\right)\ ,\\
{\rm a}^i & = & \frac{M}{r^2}\left(1\ ,0\ ,0\right)\ .
\end{eqnarray}
On the equatorial plane, $B^r=E^\theta=0$ due to symmetry, and
according to Eqs.~(\ref{Fmunu}) \& (\ref{ideal}),
\begin{equation}
E^r=\alpha r^2 {\rm v}^\phi B^\theta\ ,\ \
E^\theta=-\frac{1}{\alpha}{\rm v}^\phi B^r\ ,
\end{equation}
with $B^r_{,\theta}\neq 0$ and $E^\theta_{,\theta}\neq 0$. The
$r$-component of the last one of Eqs.~(\ref{k21}) now yields
\[
\left(1-\frac{2M}{r}\right)\frac{(r {\rm v}^\phi)^2}{r} -
\frac{M}{r^2}\nonumber
\]
\[
 - \frac{(rB^\theta)}{4\pi\rho r}
\left\{\alpha^2\left[r(rB^\theta)\right]_{,r}-B^r_{,\theta}\right\}
\left[1-(r{\rm v}^\phi)^2\right]^2 \nonumber
\]
\[
-\frac{(rB^\theta)^2}{4\pi \rho r}\left\{\frac{M}{r}
-\alpha^2(r{\rm v}^\phi)^2 \left[1+\frac{r(r{\rm
v}^\phi)_{,r}}{(r{\rm v}^\phi)}\right]\right\}\left[1-(r{\rm
v}^\phi)^2\right]\nonumber
\]
\begin{equation}
=0\ . \label{fb1}
\end{equation}
Eq.~(\ref{fb1}) is the generalization of the Newtonian radial
force-balance in Schwarzschild geometry. Given the distributions
of $\tilde{B}$ and $\rho$ with $r$ and $\theta$ in the disk, it
yields the distribution of the disk azimuthal angular velocity
${\rm v}^\phi$. Those distributions can only be obtained through
complex general relativistic magneto-hydrodynamic simulations. In
this present work, however, we are only interested in determining
the position of the innermost accretion disk layer where the PRCB
is mostly active. We will thus address this problem analytically.

The vertical thickness $h$ of the inner disk is a major uncertain
parameter since it depends on several physical parameters (thermal
pressure, radiation pressure from scattered photons, magnetic
pressure), and is the subject of ongoing investigation. According
to Fig.~1, we can approximate the value of the radial component of
the magnetic field on the upper surface of the inner disk,
$B^r|_{+h}$, with the value of the vertical component of the
magnetic field on the disk midplane. Note that $B^r=0$ on the disk
midplane. In other words,
\begin{equation}\label{B1,2}
B^r|_{+h}=-B^r|_{-h}\approx -rB^\theta\ \mbox{and}\ \
B^r_{,\theta}\approx \frac{r^2B^\theta}{h}\equiv \lambda
rB^\theta\ .
\end{equation}
We have introduced here the parameter $\lambda\equiv r/h$.
Furthermore, according to Contopoulos \& Kazanas~(1998), the
dipole magnetic field originates in the innermost optically thin
disk layer of radial extent $\delta$ such that
\begin{equation}\label{delta}
\delta \sigma_T\frac{\rho}{m_p}\approx 1\ ,
\end{equation}
in which photons coming from the central accretion disk region
penetrate. This is where the PR azimuthal electric current is
generated. Beyond that layer, the magnetic field reverses polarity
(see Fig.~1). Here $\sigma_T$ is the Thompson cross section of
photons scattered by the inner disk plasma orbiting electrons, and
$m_p$ is the proton mass. Contopoulos \& Kazanas~(1998) showed
that, in order for the PRCB battery to operate secularly (i.e.
continuously), thus generating a large scale dipolar magnetic
field interior to the disk inner edge, the reverse polarity
magnetic field needs to diffuse outward through the accreting
plasma. This occurs naturally in a turbulent accretion disk with
magnetic Prandtl number $P_m\leq 1$. We will here assume that
\begin{figure}
  \includegraphics[width=9cm]{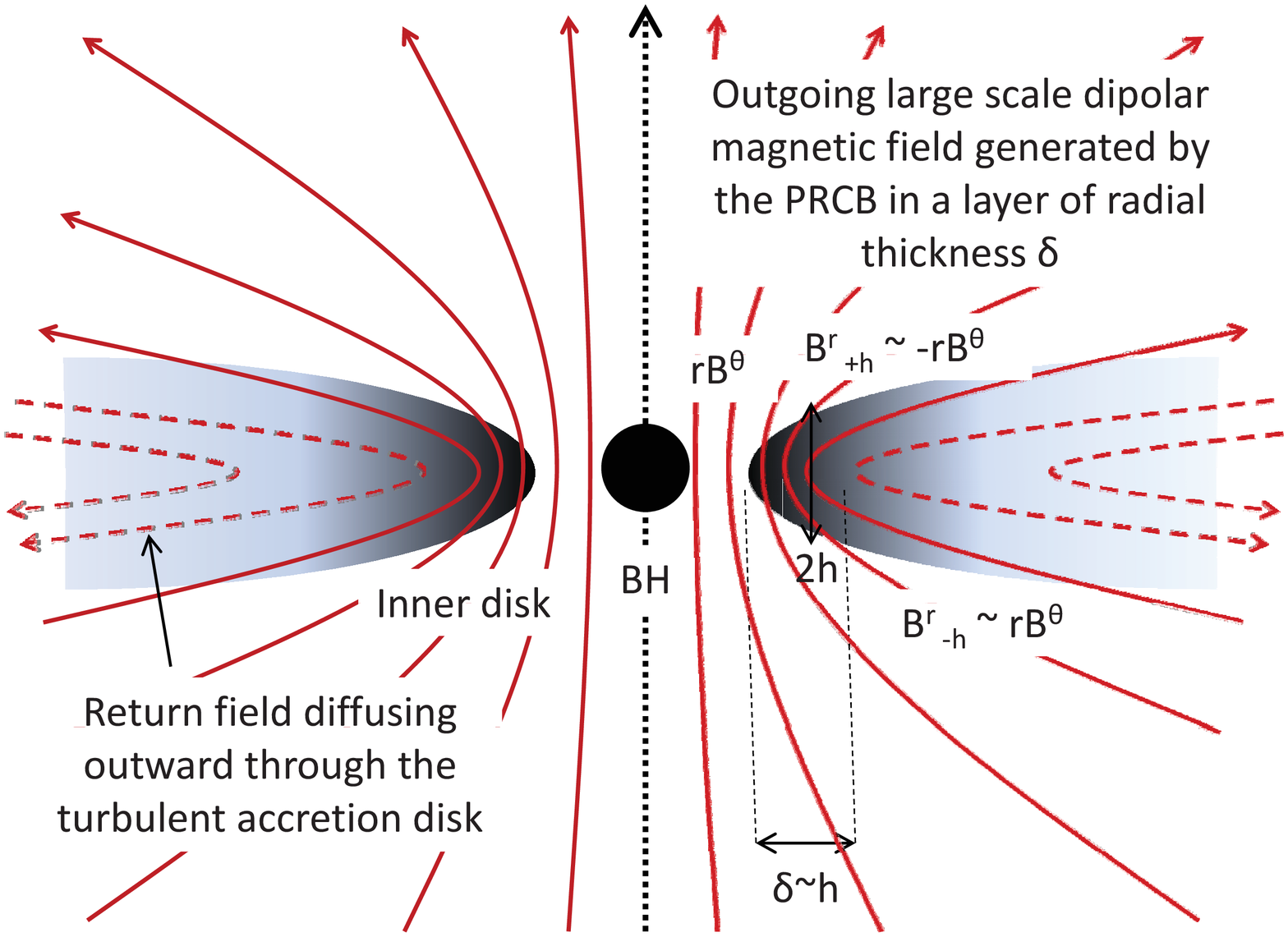}
  \caption{Sketch of innermost disk region where the large scale
dipolar magnetic field is generated by the PRCB. Solid/dashed
lines: outgoing/return field respectively. Dotted line: $z$-axis.
The PRCB layer is the innermost disk region of radial extent
$\delta$.}
 \label{landfig}
\end{figure}
\begin{equation}\label{h}
\delta\sim h\ ,
\end{equation}
although this may be generalized. Therefore, in the PR layer,
\begin{equation}\label{B2,1}
(rB^\theta)_{,r}\approx -\frac{rB^\theta}{h}\equiv -\lambda
B^\theta\ .
\end{equation}
Finally, in order to get rid of the radial derivative of ${\rm
v}^\phi$, we will assume a Keplerian velocity profile, namely
\begin{equation}\label{vKeplerian}
(r{\rm v}^\phi)_{,r} \approx -\frac{{\rm v}^\phi}{2}\ .
\end{equation}
Under the above assumptions, eq.~(\ref{fb1}) becomes
\[
\left(1-\frac{2M}{r}\right)\frac{(r {\rm v}^\phi)^2}{r} -
\frac{M}{r^2}\nonumber
\]
\[
+ \frac{{\rm v}_A^2(r)}{r} \left\{(2\lambda
-1)\left(1-\frac{M}{r}\right)\right.\nonumber
\]
\[
\left. -(r{\rm v}^\phi)^2 \left(2\lambda-\frac{3}{2}
-\frac{2M}{r}(\lambda-\frac{3}{2})\right)\right\} \left[1-(r{\rm
v}^\phi)^2\right] \nonumber
\]
\begin{equation}=0\ ,\label{c2}
\end{equation}
where, ${\rm v}_A^2(r)\equiv (rB^\theta)^2/4\pi\rho$.

In order to determine the position of the disk inner edge, we will
consider a virtual inward/outward displacement of the innermost
layer of material of radial and vertical thickness $\delta\approx
h$ and $h$ respectively, and study its stability. During that
displacement, mass conservation implies that
\begin{equation}
2\pi r \delta h \rho = \mbox{constant}\ .
\end{equation}
In order to proceed, we need to make certain further assumptions
about how the various physical quantities in our problem vary
during the above displacement. Our first assumption is that the
angular momentum parallel to the symmetry axis per unit energy $l$
is conserved, i.e. that
\begin{equation}\label{l}
\alpha {\rm v}^\phi\equiv\frac{{\rm d} \phi}{{\rm d}
t}=l\frac{\alpha^2}{r^2}\ .
\end{equation}
Our second assumption is that of flux freezing during the
displacement. This implies that magnetic flux per unit mass is
also conserved, i.e.
\begin{equation}
\frac{rB^\theta}{\rho h} = \mbox{constant.} \label{fluxfreezing}
\end{equation}
Our final assumption is that
\begin{equation}\label{hr}
\lambda\equiv \frac{r}{h} = \mbox{constant}
\end{equation}
during the infinitesimal displacement. We see that under these
assumptions, ${\rm v}_A^2$ varies inversely proportionally to $r$
during the displacement away from the position of equilibrium,
namely
\begin{equation}
{\rm v}^2_A(r) \approx {\rm v}^2_A\frac{r_{\rm ISCO}}{r}\ ,
\label{vAISCO}
\end{equation}
where ${\rm v}_A^2\equiv {\rm v}_A^2(r_{\rm
ISCO})$.\footnote{Eq.~(\ref{vAISCO}) may be generalized as ${\rm
v}_A^2(r)\approx {\rm v}_A^2 (r_{\rm ISCO}/r)^\kappa$.}

As we noted before, eq.~(\ref{fb1}) is the generalization of the
Newtonian radial force-balance, which can also be seen as the
zeroing of the first radial derivative of an effective potential
$V_{\rm eff}$. Therefore, in analogy to the Newtonian case, the
displacement stability of the innermost disk layer is determined
by the sign of the second radial derivative of $V_{\rm eff}$.
Marginal stability corresponds to zero second derivative of
$V_{\rm eff}$, i.e. to zero first derivative of eq.~(\ref{fb1}).
When we substitute eqs.~(\ref{l}) \& (\ref{vAISCO}) in
eq.~(\ref{c2}), the condition for marginal stability, becomes
\[
{\cal F}(x,\tilde{l};\lambda, {\rm v}_A^2 x_{\rm ISCO})=0\ ,\ \
\mbox{and}
\]
\begin{equation}
\frac{\partial}{\partial x} {\cal F}(x,\tilde{l};\lambda, {\rm
v}_A^2 x_{\rm ISCO})=0\ ,\label{c2b}
\end{equation}
where,
\[
{\cal F}(x,\tilde{l};\lambda, {\rm v}_A^2 x_{\rm ISCO})\equiv
\left(1-\frac{2}{x}\right)^2\frac{\tilde{l}^2}{x^3}-\frac{1}{x^2}
\nonumber
\]
\[
+ \frac{{\rm v}_A^2 x_{\rm ISCO}}{x^2} \left\{(2\lambda
-1)\left(1-\frac{1}{x}\right)\right.\nonumber
\]
\begin{equation}
\left. -\frac{\tilde{l}^2}{x^2}(1-\frac{2}{x})
\left(2\lambda-\frac{3}{2} -\frac{2\lambda -3}{x}\right)\right\}
\left[1-\frac{\tilde{l}^2}{x^2}(1-\frac{2}{x})\right]\ ,\nonumber
\end{equation}
and $x\equiv r/M$, $\tilde{l}\equiv l/M$. Notice that in this
approach we do not need to derive an expression for $V_{\rm
eff}(r)$.

The position $x_{\rm ISCO}$ of the innermost marginally stable
circular orbit as a function of our parameters $\lambda$ and ${\rm
v}_A^2$ is obtained numerically by eliminating $\tilde{l}$ in the
above system of equations (see Fig.~2 for $a=0$).
\begin{figure}
\begin{center}
\includegraphics[width=14cm]{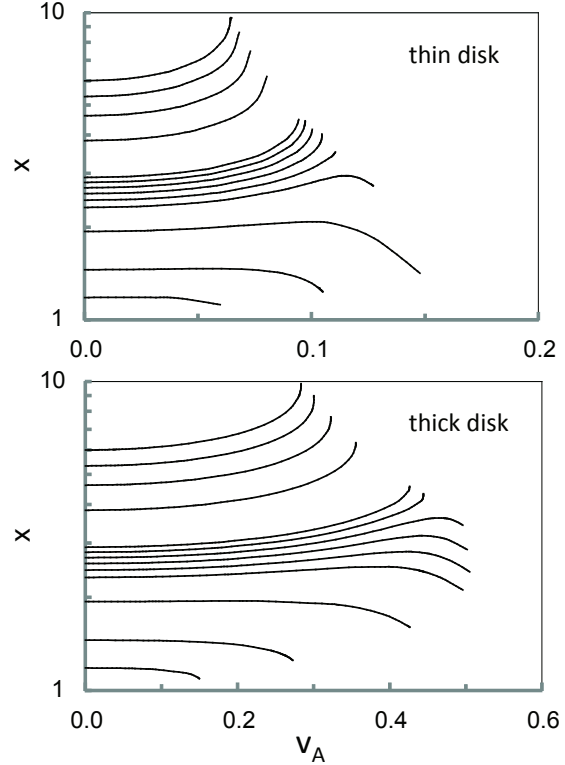}
\end{center}
\caption{Dependence of $x_{\rm ISCO}\equiv r/M$ on the inner
magnetic field strength for various black hole spin values $a$. At
the points the curves end they begin to turn up and backwards. Top
plot: thin disk ($\lambda\equiv r/h=10$). Bottom plot: thick disk
($\lambda=1$). The various curves from top to bottom correspond to
$\alpha =0, 0.2, 0.4, 0.6, 0.8, 0.82, 0.84, 0.86, 0.88, 0.9, 0.95,
0.99, 0.999M$ respectively.}
\end{figure}
For ${\rm v}_A\ll 1$, the ISCO gradually moves outward from its
unmagnetized position $6M$. In other words, {\em the magnetic
field acts to destabilize the disk inner edge}, and the disk inner
radius must be larger than $6M$ for stability, in accordance with
Lovelace {\em et al.}~(1986). However, for higher values of ${\rm
v}_A$ an ISCO ceases to exist. We conclude that, {\it in the
presence of a large scale dipolar magnetic field, the ISCO moves
outward, and eventually disappears as the field grows beyond a
threshold value}. In the case of a thin disk ($\lambda=10$), that
threshold value corresponds to ${\rm v}_A\sim 0.06$. We will
return to this important point in the Discussion section.

\section{The Kerr metric}

The above can be directly generalized in the case of a spinning
black hole. In Boyer-Lindquist coordinates, the Kerr metric reads:
\begin{eqnarray}
{\rm d}s^2 & = & g_{tt} {\rm d}t^2+2g_{t\phi}{\rm d}t {\rm d}\phi
+g_{tt} {\rm d}r^2
+g_{\theta\theta} {\rm d}\theta^2+g_{\phi\phi}
{\rm d}\phi^2\nonumber\\
& = & -(1-\frac{2Mr}{\Sigma}){\rm d}t^2-\frac{4M a r}{\Sigma}{\rm
d}t {\rm d}\phi \nonumber\\
& & +\frac{\Sigma}{\Delta} {\rm d}r^2+\Sigma {\rm
d}\theta^2+\frac{A}{\Sigma}{\rm d}\phi^2
\end{eqnarray} where
$M$ is again the mass of the black hole, $a$ its angular momentum
per unit mass $(0\leq a \leq M)$ and
\begin{equation}\label{k2}
\Delta \equiv r^2-2M r+a^2\ ,\ \ \Sigma \equiv
r^2+a^2\cos^2\theta\ ,\nonumber
\end{equation}
\begin{equation}
A \equiv (r^2+a^2)^2-a^2\Delta\ .
\end{equation}
The lapse function is equal to $\alpha=(\Delta\Sigma/A)^{1/2}$.
$\omega=2aMr/A$. On the equatorial plane ($\theta=\pi/2$),
\begin{eqnarray}
\alpha & = &
\left(1-\frac{2M}{r}+\frac{a^2}{r^2}\right)^{1/2}\frac{r^2}{\sqrt{A}}
\ ,\\
\label{Umu} U^\mu & = & \frac{1}{\alpha}\left(1\ ,0\ ,0\ ,
\frac{2aMr}{A}\right)\ ,\\
\label{ai} {\rm a}^i & = & \frac{Mr^2}{A}\left(1+2\frac{a^2}{r^2}
-4\frac{Ma^2}{r^3}+\frac{a^4}{r^4}\ ,0\ ,0\right)\ .
\end{eqnarray}
As before, $B^r=E^\theta=0$ due to symmetry,
\begin{equation}
E^r=\frac{\Delta}{\alpha} {\rm v}^\phi B^\theta\ ,\ \
E^\theta=-\frac{1}{\alpha}{\rm v}^\phi B^r\ ,
\end{equation}
with $B^r_{,\theta}\neq 0$ and $E^\theta_{,\theta}\neq 0$. In
order to estimate the position of the ISCO when the PRCB is
active, we will make the same simplifying assumptions as in the
previous section. The $r$-component of the last one of
eqs.~(\ref{k21}) simplifies then considerably and becomes
\[
\Gamma^{r}_{\phi\phi}({\rm
v}^\phi)^2+\left[\frac{1}{\alpha}(\Gamma^r_{t\phi}+\omega
\Gamma^r_{\phi\phi})+g_{\phi\phi}\sigma^{r\phi}\right]{\rm
v}^\phi+{\rm a}^r \nonumber
\]
\[
-\frac{{\rm v}_A^2 r_{\rm
ISCO}}{r^2\Gamma^4}\{\frac{\Delta}{r^2}(\lambda+1)+\lambda\}
\]
\begin{equation}
-\frac{{\rm v}_A^2 r_{\rm ISCO}}{2r^3\Gamma^2}\{X({\rm v}^\phi)^2
+Y{\rm v}^\phi+Z\} = 0\ , \label{Kerr}
\end{equation}
where the detailed expressions for $\sigma^{r\phi}$, $X$, $Y$,
$Z$, the Christoffel symbols $\Gamma^r_{t\phi}$,
$\Gamma^{r}_{\phi\phi}$ and the flow Lorentz factor $\Gamma$ can
be found in the Appendix. As before, ${\rm v}_A^2\equiv {\rm
v}_A^2(r_{\rm ISCO})$.

The assumption that angular momentum parallel to the symmetry axis
per unit energy $l$ is conserved during a virtual infinitesimal
displacement of the innermost layer of the disk is now generalized
as
\[
\alpha{\rm v}^\phi\equiv\frac{{\rm d} \phi}{{\rm d} t}-\omega
\nonumber
\]
\[
=\frac{-\Delta(a-l)+a(r^2+a^2)-la^2}{ -a(a-l)\Delta+(r^2+a^2)^2-la
(r^2+a^2)}-\omega\nonumber
\]
\begin{equation}
=l\frac{\left(1-\frac{2M}{r}\right)
\frac{r^2}{A}+\omega^2}{1-\omega l}\label{lKerr}
\end{equation}
(Bardeen, Press \& Teukolsky~1972; hereafter BPT72). When we
substitute the above in eq.~(\ref{Kerr}), we obtain a complicated
expression of the form
\begin{equation}
{\cal G}(x,\tilde{l};\lambda, {\rm v}_A^2 x_{\rm ISCO})=0\ ,
\label{first}
\end{equation}
and the condition for marginal stability becomes that
\begin{equation}
\frac{\partial}{\partial x} {\cal G}(x,\tilde{l};\lambda, {\rm
v}_A^2 x_{\rm ISCO})=0\ .\label{second}
\end{equation}
As before, $x\equiv r/M$, and $\tilde{l}\equiv l/M$. The position
of the ISCO as a function of our two free parameters $\lambda$ and
${\rm v}_A^2$ is obtained as a solution of the above system of
equations (\ref{first}) \& (\ref{second}) for $x$ and $\tilde{l}$.
The solution for various values of the black hole spin parameter
$a$ is obtained numerically in Fig.~2 for two values of
$\lambda\equiv r/h$ that correspond to a thin and thick disk
respectively. Similarly to the case of a Schwarzschild black hole,
for small values of ${\rm v}_A$, the ISCO always moves outward.
However, for values of $a\gsim 0.8M$, an interesting transition
takes place: as ${\rm v}_A$ grows beyond a threshold value, {\em
the ISCO moves back inward}. As we will now see, this result
complicates the observational determination of the black hole
spin.

\section{Discussion}

The displacement of the ISCO in the presence of a large scale
magnetic field generated in situ has direct implications in the
currently most promising method for measuring black hole spin,
namely fitting the thermal X-ray continuum (Zhang et al.~1997).
This method requires knowledge of both the Schwarzschild radius
and the radius of the inner edge of the disk, i.e. it requires
knowledge of $x_{\rm ISCO}$. As is shown in BPT72, in an
unmagnetized disk, $x_{\rm ISCO}$ is one-to-one related to the
spin of the black hole. In the present work, we showed that
$x_{\rm ISCO}$ depends also on the value of the magnetic field
accumulated in the innermost region of the accretion disk. This
result introduces one extra complication.

In particular, the disappearance of the ISCO as the field grows
towards equipartition (defined in this context as ${\rm
v}_A\rightarrow 1$) is an indication for a dramatic change in the
structure of the accretion disk in the presence of a strong enough
magnetic field generated by the PRCB. It is natural to speculate
that, when the conditions for marginal stability of the inner edge
cannot be satisfied, the accumulated magnetic field will escape
outward through magnetic Rayleigh-Taylor-type instabilities, as
proposed in Igumenshchev~(2008), Tchekhovskoy, Narayan \&
McKinney~(2011), and Kylafis~{\em et al.}~(2012). As a result, the
accumulated magnetic field will decrease, the ISCO will reappear
close to its unmagnetized disk position, and the disk will undergo
a new phase of field growth and evolution. For a stellar mass
black hole, the time required for equipartition field growth is on
the order of a few hours to a few days (Contopoulos \&
Kazanas~1998; Kylafis {\em et al.}~2012), whereas the intervals of
dramatic disk structure change are expected to be much shorter
(dynamic timescales).

For black holes with spin parameter $a< 0.8M$, the inner edge of
the disk is expected to be outside its unmagnetized position,
therefore the thermal X-ray continuum fitting method will on
average yield a {\em systematically smaller} value for the black
hole spin. On the other hand, for fast spinning black holes with
$a\gsim 0.8M$, the inner edge of the disk moves to lower values,
and in those cases the thermal X-ray continuum fitting method will
on average yield {\em systematically larger} values for the black
hole spin. This result introduces an {\em artificial segregation}
in our observational method that may be related to the seeming
absence of evidence for the black hole spin powering the jets in
X-ray binaries (Fender, Gallo \& Russell~2010; see however also
Narayan \& McClintock~2012). In other words, the introduction of
one extra free parameter, namely the amount of magnetic flux
accumulated around the inner edge of the disk, complicates the
direct determination of the black hole spin through the knowledge
of $x_{\rm ISCO}$ alone.

The above are summarized in Fig.~3 where we show the range of
$x_{\rm ISCO}$ values as a function of $a$ for a thick disk
($r/h=1$ at its inner edge). The dashed curve corresponds to the
unmagnetized disk (BPT72), whereas the solid curve to a maximally
magnetized disk that supports an ISCO. The dashed-dotted curve
corresponds to the maximum values of $x_{\rm ISCO}$ reached as the
inner field (${\rm v}_A$) grows in a spinning black hole with
$a\gsim 0.8M$. In other words, according to our simplified model
of the inner disk, observations of $x_{\rm ISCO}$ must lie in the
shaded region. As an example, in the most studied object, LMC X-3,
estimates of $x_{\rm ISCO}$ range from about 4 to about $6$
(Steiner {\em et al.}~2010, assuming a canonical value for the
black hole mass equal to $7.5M_{\odot}$). If the PRCB did not
operate and the disk were unmagnetized, these values would yield
an average black hole spin value $a\sim 0.3M$ (McClintock~{\em et
al.}~2011). It is natural, though, to expect that the PRCB does
operate as described above, and that the various estimates of
$x_{\rm ISCO}$ simply correspond to the various stages of magnetic
field growth that the source is found to be in. As shown in
Fig.~3, in this context, a more natural value of the black hole
spin in LMC X-3 would then be $a\sim 0.6M$.
\begin{figure}
  \includegraphics[width=10cm]{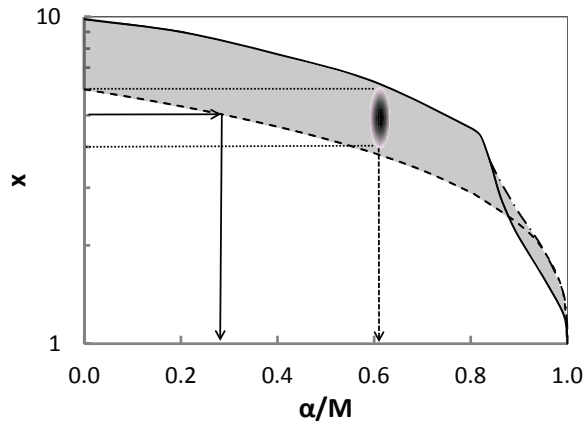}
  \caption{Range of $x_{\rm ISCO}$ values as a function
  of the black hole spin parameter $a$ for a thick disk $r/h=1$.
  Dashed curve: unmagnetized disk. Solid curve: maximally magnetized
  inner disk that supports an ISCO. Shadded region: range of
  $x_{\rm ISCO}$ values when $0\leq {\rm v}_A\leq \left.{\rm
  v}_A\right|_{\rm max}$. Diffuse cloud: range of
  $x_{\rm ISCO}$ values obtained for the X-ray black hole binary
  LMC X-3 over a period of 25 years. Left vertical arrow:
  average spin parameter of unmagnetized disk. Right vertical arrow:
  spin parameter that yields the observed range of $x_{\rm ISCO}$
  values in the context of the PRCB.}
 \label{spin}
\end{figure}

Finally, the reader may wonder in what respect the PRCB manifests
itself in the inner disks structure differently from flux
advection from large distances. One way to differentiate between
the two is the radial width $\delta$ of the PRCB layer over which
$B^\theta$ changes significantly. In the PRCB and a thin disk, our
parameter $r/\delta\sim \lambda$ is expected to be on the order of
10. On the other hand, in the PRCB and a thick disk, as well as in
a disk where the magnetic field is advected inward from large
distance, $\lambda$ is expected to be on the order of unity.
Another way to differentiate between the two possibilities has to
do with the particular cyclic disk variability central to the
PRCB, namely intervals of secular field growth (hours, days)
separated by intervals of strong (msec) disk variability. Our
present investigation on the role of the PRCB in determining the
structure of the inner disk is certainly not exhaustive and will
continue in the framework of Program ARISTEIA of the General
Secretariat for Research and Technology of Greece.

\section*{Acknowledgments}

We thank Drs. Dimitris Christodoulou and Demos Kazanas for their
comments and suggestions. D.B.P. would like to thank the Institute
fur Astronomie und Astrophysik Abteilung Theoretische Astrophysik
of Eberhard Karls Universitat Tubingen where part of this work was
performed.

{}

\appendix

\section[]{}

Various expressions used throughout the paper are defined below
according to Thorne \& Macdonald~(1982):
\[
\tilde{\nabla}\cdot \tilde{A} \equiv A^{j}_{;j}\ ,\ \
(\tilde{\nabla}\times \tilde{A})^i \equiv \epsilon^{ijk}
A_{k;j}\nonumber
\]
\[
D_{\tau}A^{\beta} \equiv A^{\beta}_{;\mu}U^{\mu}-U^{\beta}U_{\mu ;
\nu}A^{\mu}U^\nu\nonumber
\]
\[
\tilde{A}\cdot\tilde{B} \equiv A^i B_i\ ,\ \  (\tilde{A}\times
\tilde{B})^i  \equiv \epsilon^{ijk} A_{j}B_{k}\ ,\ \
(\tilde{A}\otimes\tilde{B})^{ij}\equiv A^i B^j
\]
Here,  $\epsilon^{ijk}\equiv [ijk]/\sqrt{|{\rm
det}(\gamma_{ij})|}$ is the spatial Levi-Civita tensor. On the
equatorial plane $(\theta=\pi/2)$,
\[
\Delta=r^2-2Mr+a^2\ ,~~\Sigma=r^2\ ,A=r^4+a^2 r^2 +2Ma^2 r
\nonumber
\]
\[
\Gamma=\left[1-\frac{A}{r^4}(r{\rm v}^\phi)^2\right]^{-1/2}
\]
\[
\Gamma_{t\phi}^r = -\frac{Ma\Delta}{r^4}\ ,~~\Gamma_{rr}^r
 = -\frac{Mr-a^2}{\Delta r}\nonumber
\]
\[
\Gamma_{\theta\theta}^r = -\frac{\Delta}{r}\ ,~~
\Gamma_{\phi\phi}^r = -\frac{\Delta}{r}
\left[1-\frac{Ma^2}{r^3}\right]
\]
\begin{eqnarray}
\Gamma_{k1}^k & \equiv &\Gamma_{rr}^r+\Gamma_{\theta
r}^\theta+\Gamma_{\phi r}^\phi
\nonumber\\
& = & \frac{r}{\Delta}\left(2-\frac{5M}{r}
+\frac{2a^2}{r^2}-\frac{Ma^2}{r^3}\right)\nonumber
\end{eqnarray}
\[
\sigma^{r\phi}=\sigma^{\phi r}=
-\frac{Ma\Delta(a^2+3r^2)}{rA\sqrt{A\Delta}}\
\]
\[
{\rm a}^r=\frac{Mr^2}{A} \left(1+2\frac{a^2}{r^2}
-4\frac{Ma^2}{r^3}+\frac{a^4}{r^4}\right)\nonumber
\]
\[
\omega=\frac{2Mar}{A},~~\alpha=\sqrt{\frac{\Delta r^2}{A}}
\]
\begin{eqnarray}
X & \equiv & \left(\frac{\Delta}{\alpha}\right)^2 \{-\frac{{\rm
a}^r r^2}{\Delta}+Q-\frac{1}{\Delta}
\Gamma_{\theta\theta}^r-\frac{1}{\Delta}\Gamma_{\phi\phi}^r
(\gamma^{\phi\phi}r^2)\} \nonumber\\ & &
+2\left(\frac{\Delta}{\alpha}\right)
\left(\frac{\Delta}{\alpha}\right)_{,r} +2\frac{{\rm
v}^\phi_{,r}}{{\rm v}^\phi}\frac{\Delta^2}{\alpha^2}
\nonumber\\
Y&\equiv
&-2\frac{r^2}{\alpha}(\Gamma_{t\phi}^r+\omega\Gamma_{\phi\phi}^r)]
\nonumber\\
Z&\equiv &-{\rm a}^r r^2 -\Delta Q-\Delta_{,r}
+\Gamma_{\theta\theta}^r-\Gamma_{\phi\phi}^r
(\gamma^{\phi\phi}r^2)\nonumber\\
Q&=& \frac{\alpha_{,r}}{\alpha}+2\Gamma_{rr}^r+\Gamma_{\theta
r}^\theta+\Gamma_{\phi r}^\phi\nonumber
\end{eqnarray}

\label{lastpage}

\end{document}